\documentclass[
    ,final            
  ]
  {aipproc}

\layoutstyle{6x9}
\usepackage{color}

\begin{document}

\title{DIS2011 Heavy Flavours Session Summary 
(WG5)\footnote{Presented at the 
{\sl XIX International Workshop on Deep-Inelastic Scattering and Related Subjects 
\hbox{\bf (DIS 2011)}}, 
April 11-15, 2011, 
Newport News, VA USA}}


\author{Olaf Behnke}{
  address={DESY, D-22607 Hamburg, Germany },
}

\author{Alan Dion}{
  address={BNL},
}

\author{Fredrick Olness}{
  address={Southern Methodist University, Dallas, TX 75275, USA},
}

\begin{abstract}
We summarize the presentations of the Heavy Flavours working 
group for the 2011 DIS Workshop. 
This session contained presentations on
theoretical methods and experimental measurements of heavy quark
production, and the impact on recent experimental results from HERA,
RHIC, Tevatron, and LHC.

\end{abstract}

 \maketitle

%
%
%
%
\section{Overview}

The production of heavy quarks in high energy processes has become
an increasingly important subject of study both theoretically and
experimentally. 
The theory of heavy quark production in perturbative
Quantum Chromodynamics (PQCD) is more challenging than that of light
parton (jet) production because of the new physics issues brought
about by the additional heavy quark mass scale. The correct theory
must properly take into account the changing role of the heavy quark
over the full kinematic range of the relevant process from the threshold
region (where the quark behaves like a typical {}``heavy particle'')
to the asymptotic region (where the same quark behaves effectively
like a parton, similar to the well known light quarks $\{u,d,s\}$).

The experimental measurements of the heavy quarks are also challenging as 
the heavy quark signal must often be extracted from underneath the 
dominant light quark process. Nevertheless, the range and precision of
the heavy flavor measurements contines to improve and enable 
incisive tests of the various production mechanisms.


\section{Theoretical Challenges }

\begin{figure}
  \includegraphics[width=.30\textwidth]{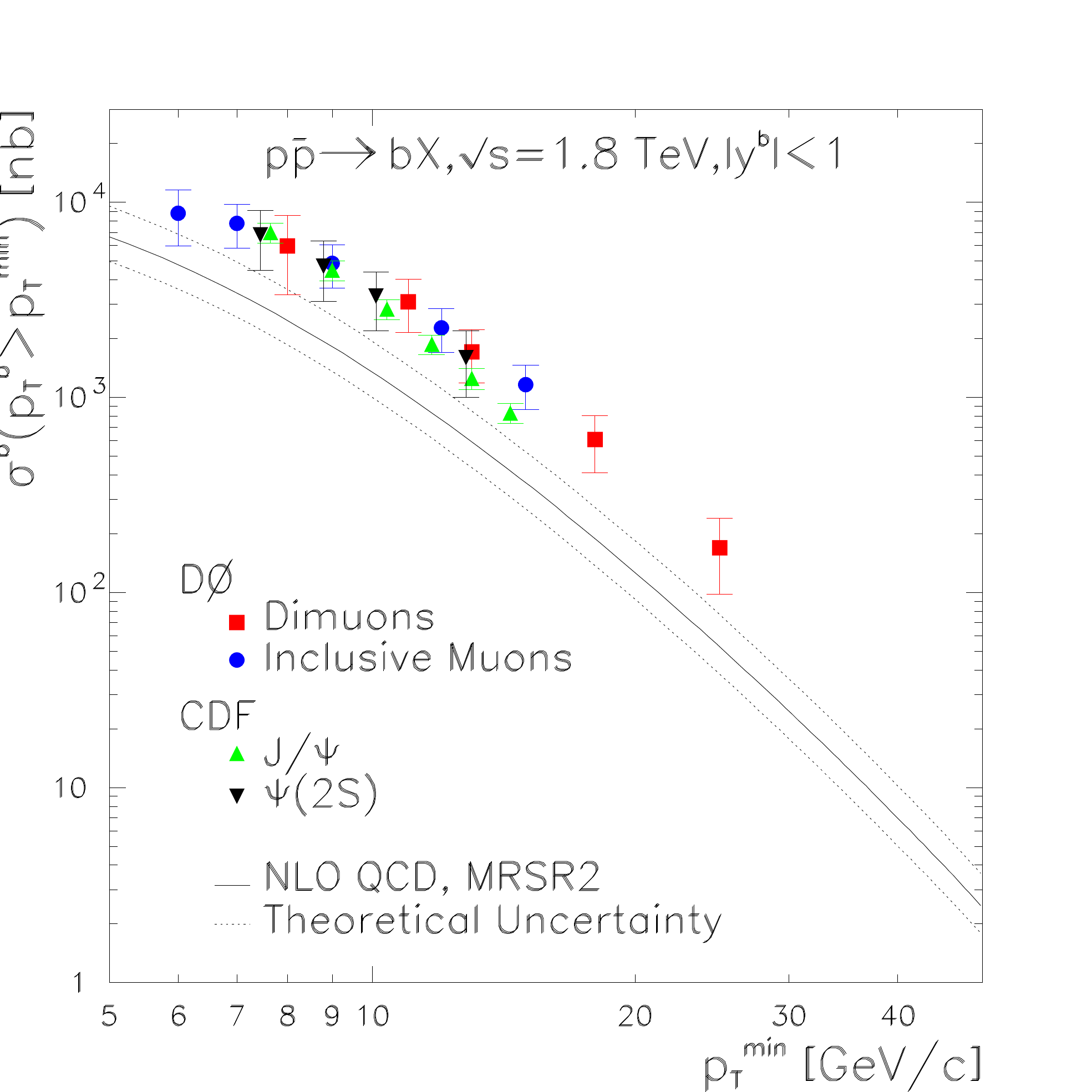}
  \includegraphics[width=.30\textwidth]{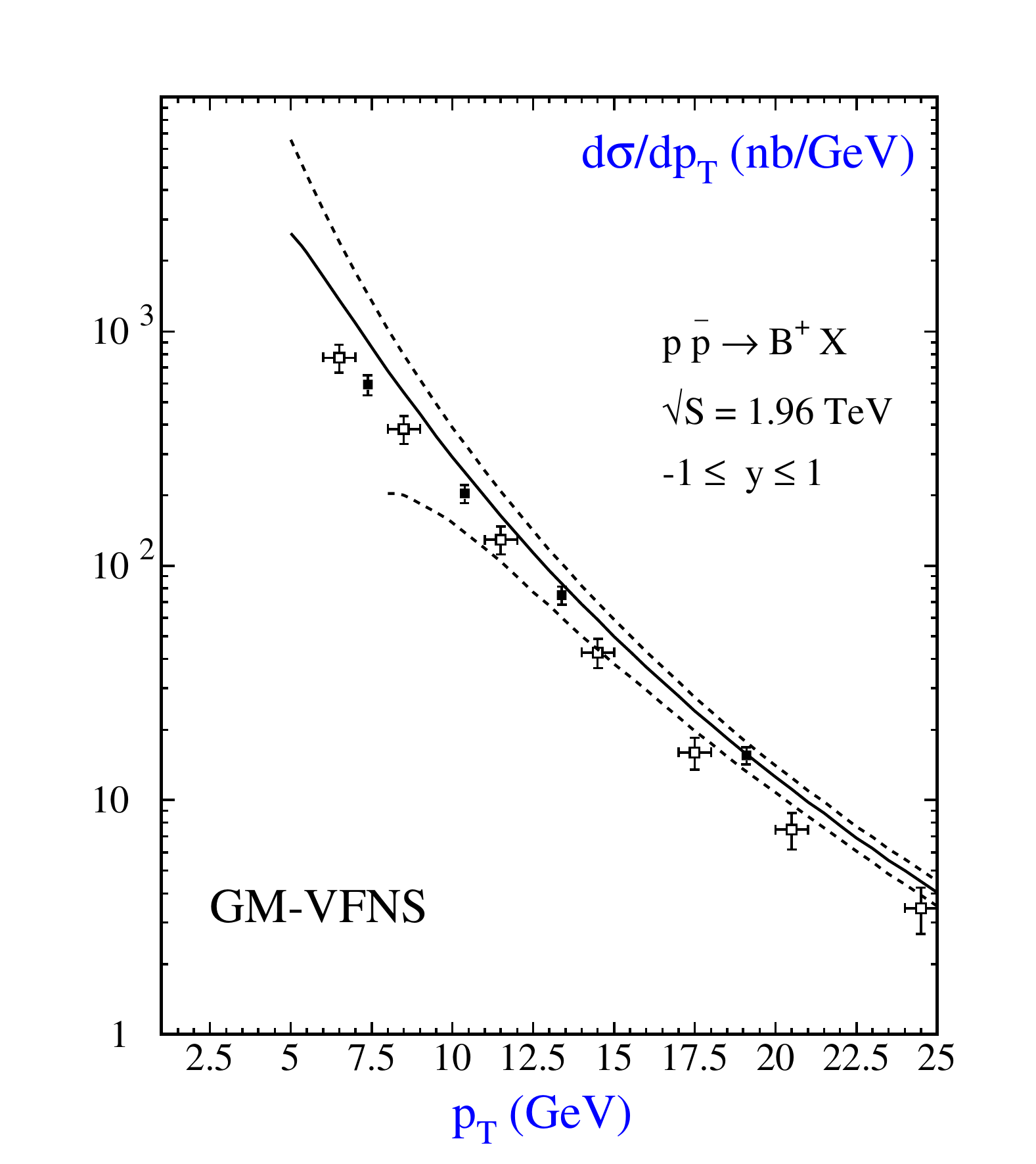}
  \includegraphics[width=.30\textwidth]{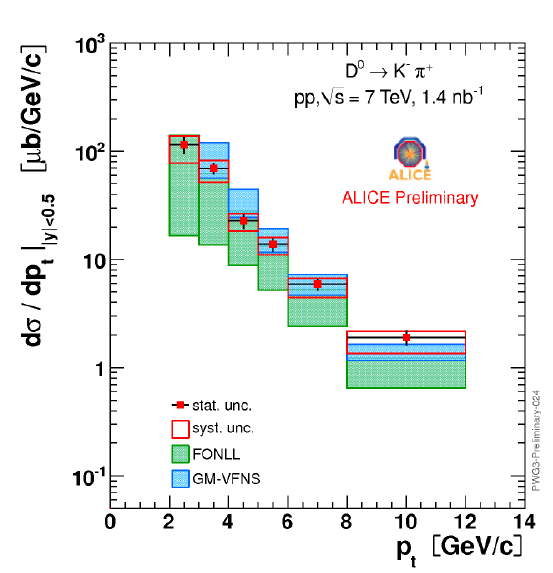}
  \caption{a) The status of b-quark hadroproduction circa 1999. 
b) Updated GM-VFNS analysis of B hadroproduction circa 2011.
c) Preliminary D hadroproduction data from the LHC.  \label{fig:bquark}}
\end{figure}

Large momentum transfer production of particles and jets provided some
of the early tests of QCD. Indeed, by taking into account quark and
gluon initiated scattering processes, QCD was able to explain the
relatively large cross sections observed at the CERN ISR. However, as
the precision of the experiments increased, there was a corresponding
need for increased theoretical precision and discrepancies arose
between theory and data for several large momentum transfer processes.

One such long-standing issue in QCD was the calculation of the cross
section for the production of b-quarks at high energy hadron
colliders. In the 1999 CERN LHC Workshop proceedings (hep-ph/0003142),
the integrated transverse momentum spectrum for b production at the
Tevatron vs. $p_T$ was a factor of 2 above the NLO theoretical
predictions (c.f., Figure~\ref{fig:bquark}-a).  
The resolution of this discrepancy was multifaceted
involving improvements on both the experimental and theoretical side;
as a result of these improvements, we are now able to find good
agreement between the data and theory for these observables.


\begin{figure}
  \includegraphics[width=.24\textwidth]{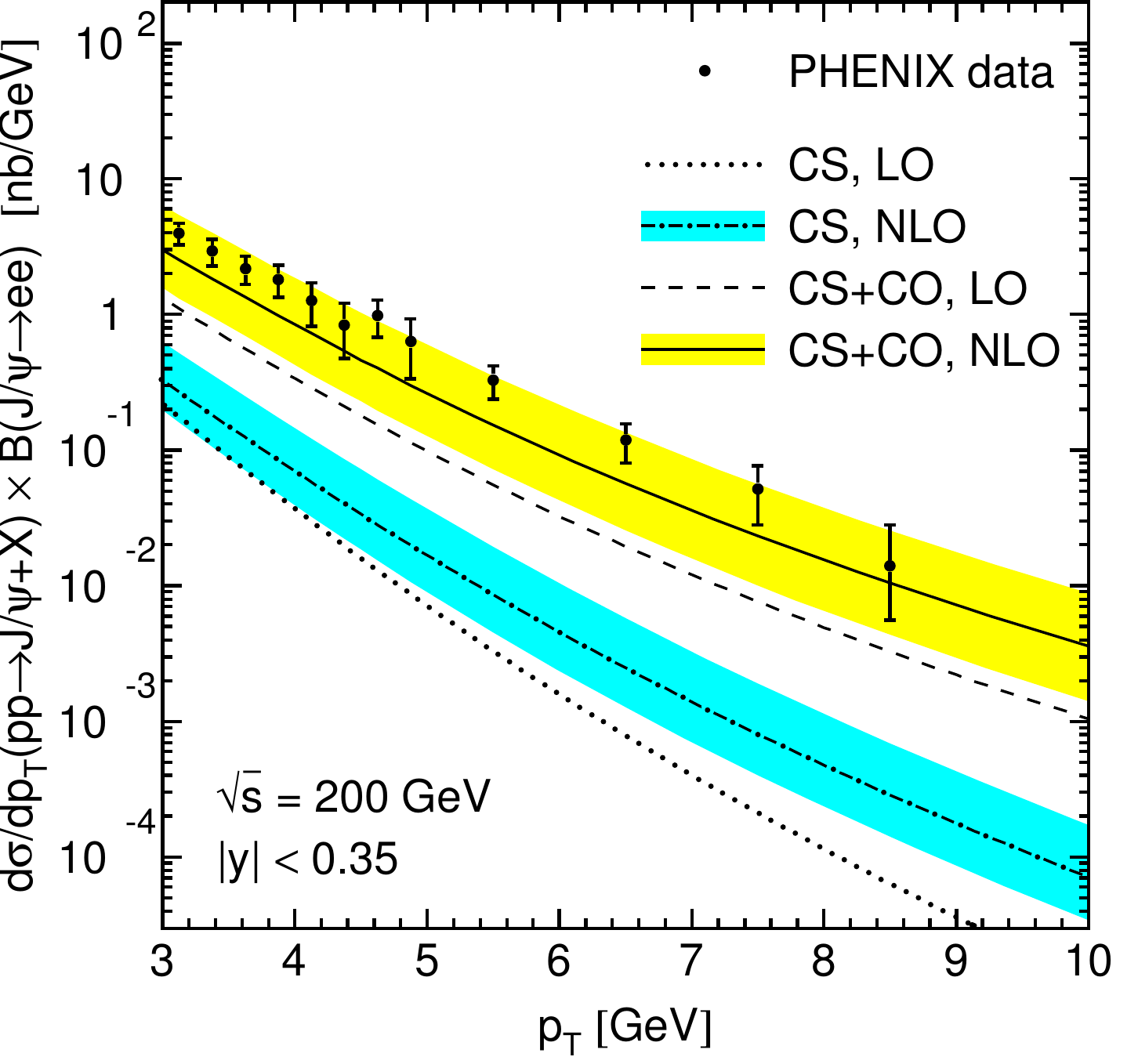}
  \includegraphics[width=.24\textwidth]{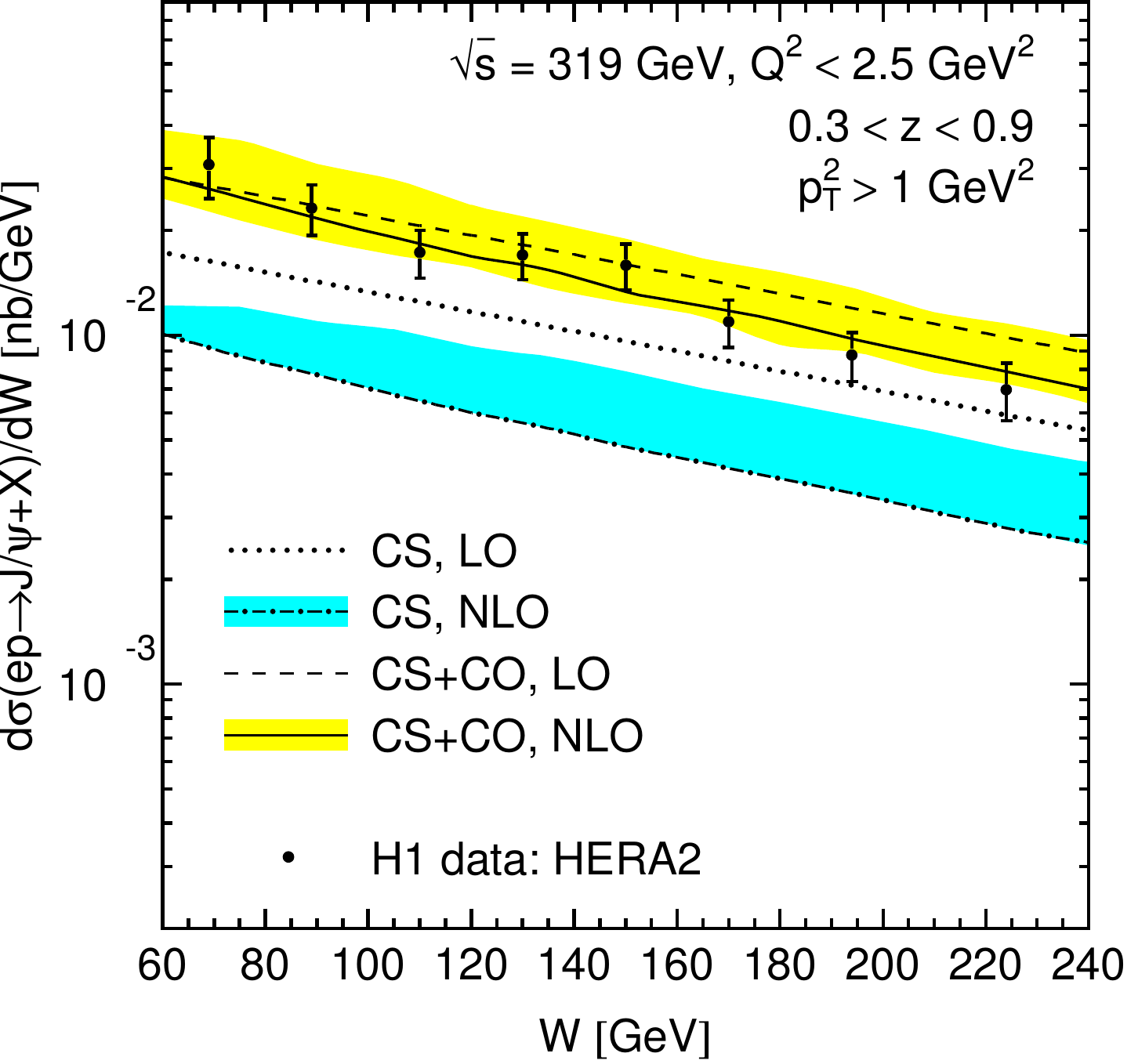}
  \includegraphics[width=.24\textwidth]{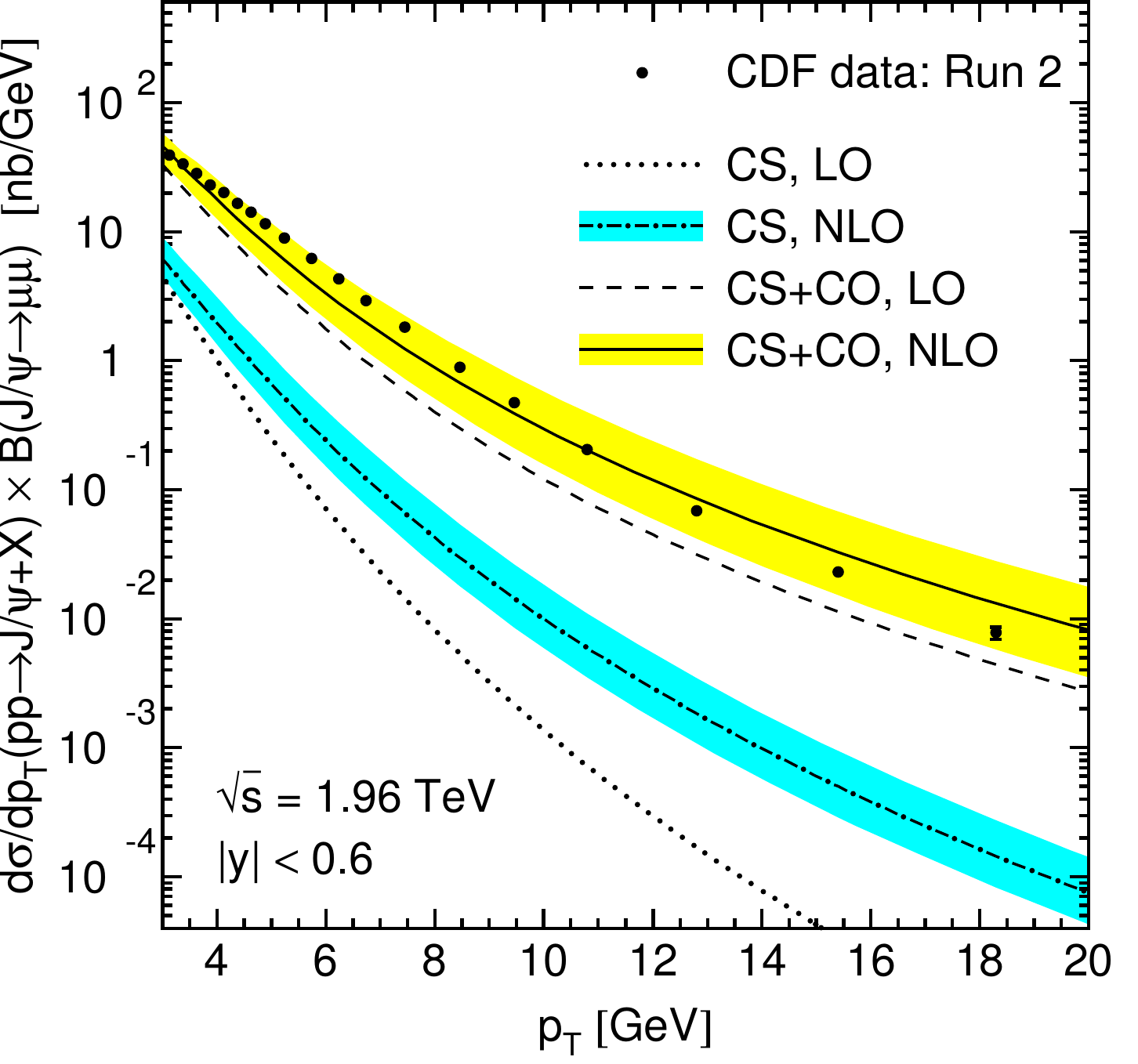}
  \includegraphics[width=.24\textwidth]{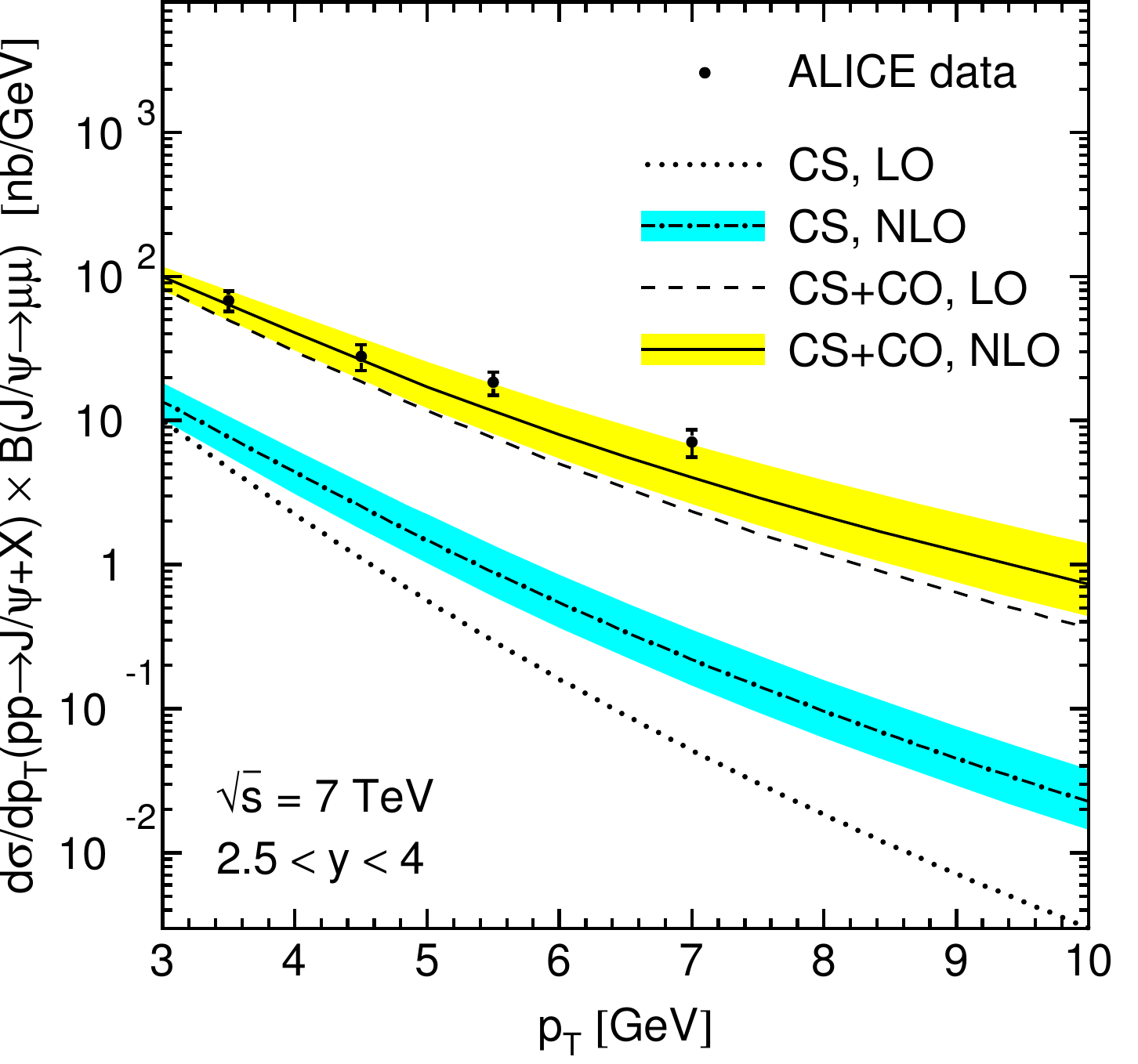}
  \caption{$J/\Psi$ production for a variety of experiments
compared with the Color Singlet (CS), Color Octet (CO), Leading Order (LO),
and Next-to-Leading Order (NLO) predictions. \label{fig:quarkonia}}
\end{figure}

Bernd Kniehl (in collaboration with G. Kramer, I. Schienbein, and
H. Spiesberger) presented updated results on the hadro-production of
D and B mesons.\cite{kniehl1} 
They used the General-Mass Variable Flavor Number
Scheme (GM-VFNS) calculation with updated fragmentation functions
extracted from LEP and B-factory data.  Figure~\ref{fig:bquark}-b)
displays the results for B meson production at the Tevatron, and
Figure~\ref{fig:bquark}-c) displays the results for D meson production
at the LHC.  The GM-VFNS employs the full mass dependence in the
calculation, and the fragmentation functions include the full scaling
violations.  The good agreement observed between the data and theory
validates this approach and 
reaffirms our ability to make accurate calculation for 
all mass scales---from small to large---with this parameter-free formalism.


In a separate presentation, Bernd Kniehl (in collaboration with
M. Butenschon) discussed recent calculations of the cross sections for
inclusive $J/\psi$ production at next-to-leading order (NLO) within
the factorization formalism of non-relativistic quantum chromodynamics
(NRQCD), including the full relativistic corrections due to the
intermediate $^1\!S_0^{[8]}$, $^3\!S_1^{[8]}$, and $^3\!P_J^{[8]}$
color-octet states.\cite{kniehl2} 
In this context, they performed a NLO global fit of the respective
color-octet long-distance matrix elements to all available
high-quality data of inclusive unpolarized $J/\psi$ production,
including KEKB, LEP~II, RHIC, HERA, the Tevatron, and the LHC,
comprising a total of 194 data points from 26 data sets.
Selected results are displayed in Figure~\ref{fig:quarkonia}.  The fit
values of the color-octet (CO) long-distance matrix elements (LDMEs)
agree with their previous analysis, demonstrate the validity of NRQCD
factorization for charmonium, and provides evidence for LDME
universality and the existence of CO processes.  Future work on
polarized $J/\psi$ production could provide additional information on
these terms.


\begin{figure}
  \includegraphics[width=0.40\textwidth]{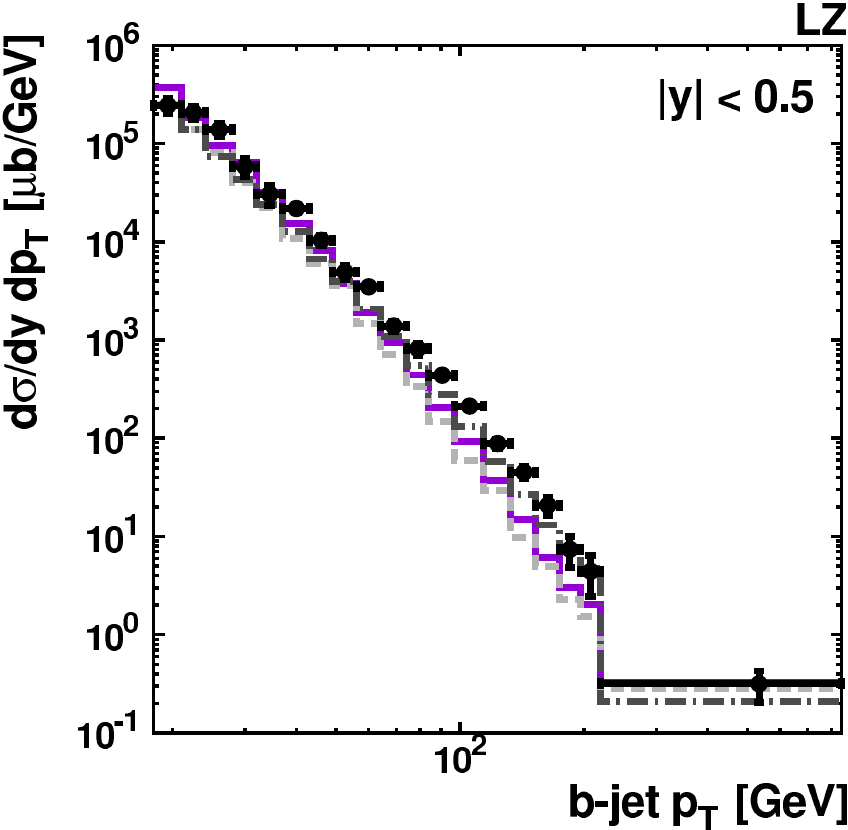}
\qquad
  \includegraphics[width=0.40\textwidth]{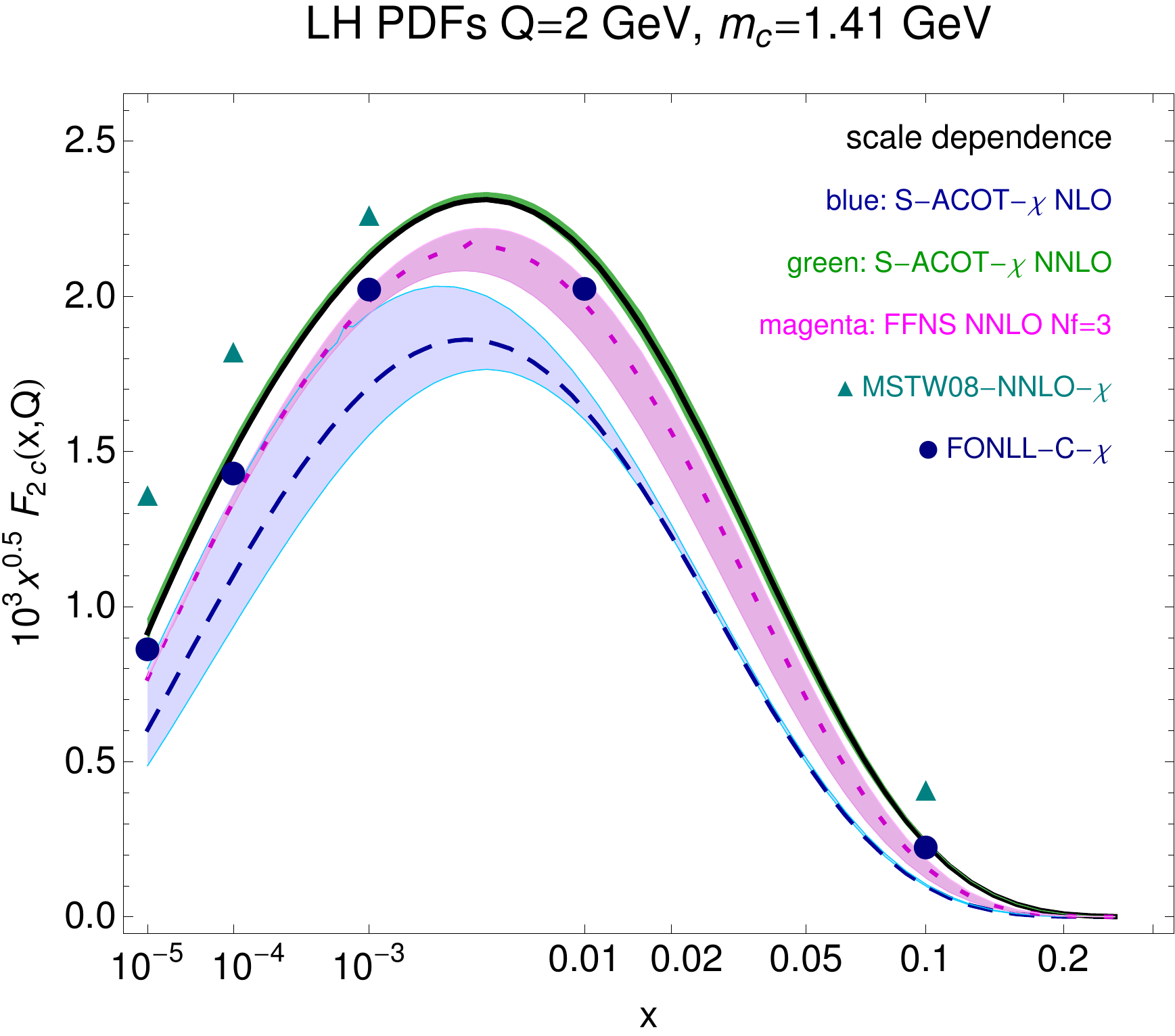}
  \caption{%
a) b-jet $p_T$ differential cross section computed with the 
CASCADE Monte Carlo generator. 
b) $F_2^{c}$ comparisons at NNLO. 
\label{fig:zotov}}
\end{figure}

N.P.~Zotov (in collaboration with H. Jung, M. Kraemer, A.V. Lipatov)
presented predictions from the CASCADE Monte Carlo generator for b-jet
and quarkonium production at the LHC.\cite{zotov} 
CASCADE  uses the $k_T$-factorization approach with unintegrated
($k_T$-dependent) gluon distribution (UGD).  The UGDs in a proton are
determined using the CCFM evolution equation and the
Kimber-Martin-Ryskin (KMR) prescription.
With conventional PDFs, the exchanged parton is assumed to be
on-shell, and collinear with zero $k_T$. In contrast, the UGD allows
for non-zero $k_T$ contributions even at Leading-Order (LO); hence it has
the potential to provide an improved description of the physics at a
given order in perturbation theory.
Results for inclusive b-jet and quarkonium production have been
computed in this approach for the CERN LHC energies, and a sample
comparison with the CMS data is shown in Figure~\ref{fig:zotov}-a).
The overall description of the data is reasonable, and in most cases
it is comparable to the standard Monte Carlo calculation; notably,
there are some distributions where the $k_T$-factorization approach
yields an improved description of the data.


Christian Pascaud presented a new ``Continuous Flavor Number Scheme''
(CFNS) for heavy flavors in DIS.\cite{pascaud} 
The goal was to modify the DGLAP
evolution to include the heavy quarks at all scales and suppress the
heavy flavors at high x in the low Q region based upon the kinematical
constraints (i.e., $W>2M_h$), the momentum sum-rule, and the quark
counting sum-rules.  Thus, for $Q^2 << 4 M^2_h$ the corresponding
heavy quark is decoupled, while for $Q^2 >> 4 M^2_h$ the DGLAP
equations and coefficient functions regain their usual
appearance. Comparisons with the conventional schemes were made for
both the fitted PDFs as well as the $F_2^{c\bar{c}}$ structure
function.


\begin{figure}
  \includegraphics[width=0.45\textwidth]{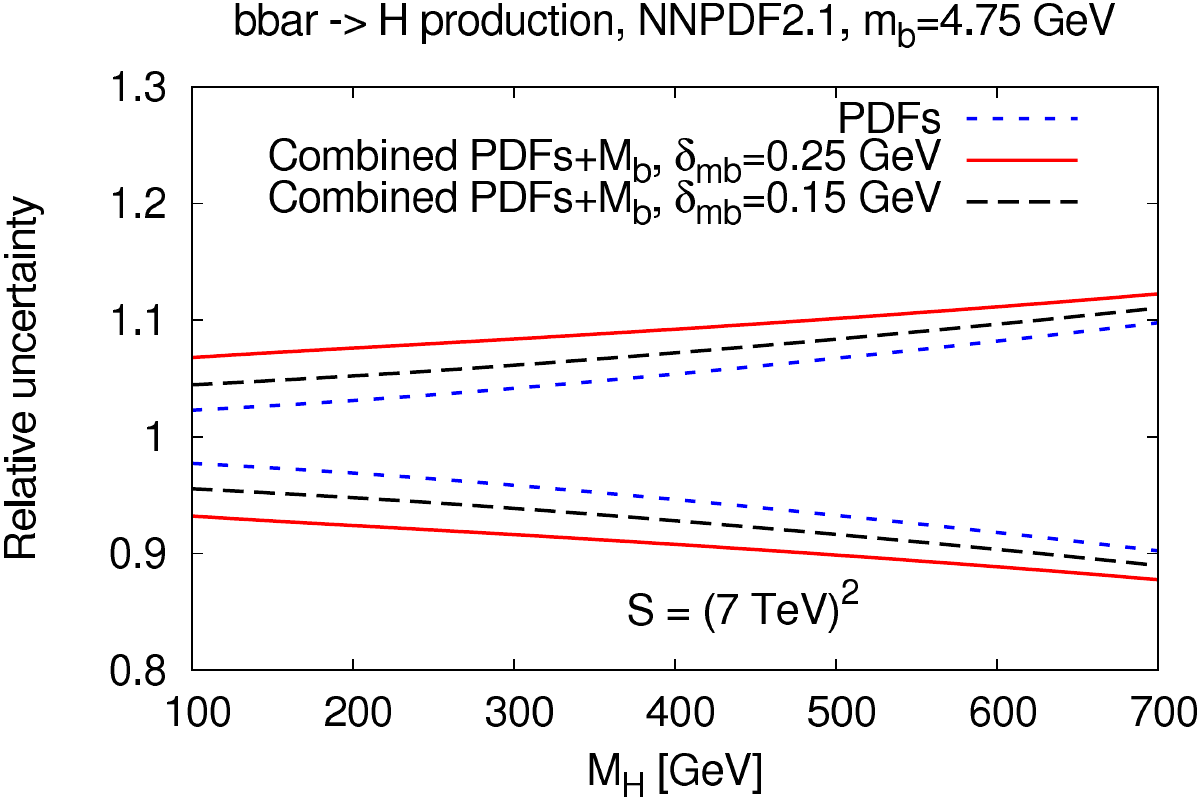}
\qquad
  \includegraphics[width=0.35\textwidth]{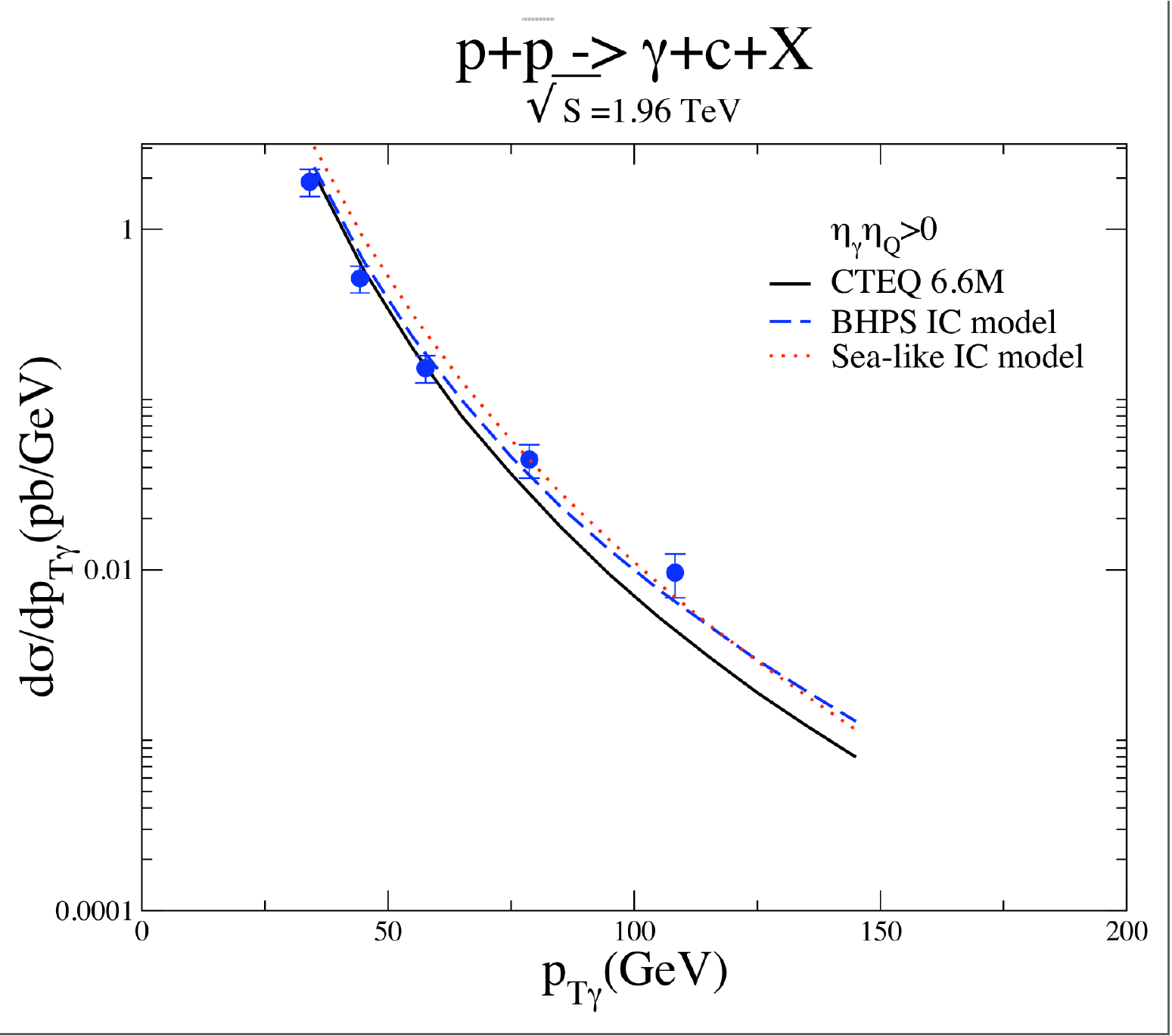}
  \caption{%
a) Combined PDF+$m_b$ uncertainties ($1\sigma$) on the total 
cross section for $b\bar{b}\to H$ production. 
b) The differential cross section for $p\bar{p} \to \gamma c X$
using the CTEQ6.6M PDFs, and two Intrinsic Charm (IC) models. 
\label{fig:rojoTZ}}
\end{figure}

Nikolaos Kidonakis presented new results on the top quark cross
section and differential distributions.\cite{kidonakis}
Specifically,
results for $t\bar{t}$ and single top production channels performed at
NNLL resummation were presented. 
The resummed NNLL calculation results in reduced scale dependence of
the cross section as compared to the lower order
results. Additionally, the enhancements due to the NNLO approximation
are significant as compared to the NLO results.
For the $t\bar{t}$ channel at the LHC the enhancement is +7.6\% at
1$\sqrt{s}=7$~TeV and +8.0\% at $\sqrt{s}=14$~TeV.
Similarly, 
For single top at the LHC the shift is -1\%
to -3\% in the $t$-channel, and +13\% in the $s$-channel.
These processes are  precision ``standard candle''
measurements for the LHC and also can serve to
calibrate the Higgs search.


F. Wissbrock (in collaboration with J. Ablinger, J. Blumlein,
S. Klein, and C. Schneider) presented the first results for 3-Loop
Heavy Flavor Corrections to DIS with two Massive Fermion
Lines.\cite{Ablinger:2011pb}
Specifically, these are the contributions at order ${\cal
O}(\alpha_S^3)$ which are proportional to $T_F^2 C_{F,A}$.  This
calculation is part of a broader program to extend the analysis of the
PDFs to NNLO order while retaining the massive contributions.


M. Guzzi (in collaboration with P. Nadolsky, C.-P. Yuan, and H.L. Lai)
presented preliminary results on the S-ACOT-$\chi$ heavy flavor
contributions at NNLO.\cite{guzzi} 
This also is a key ingredient for extending the
PDF analysis to higher order in the perturbation theory. Preliminary
results (sample shown in Figure~\ref{fig:zotov}-b) indicates that
this approach for the NNLO calculation of $F_2^{c,b}$ and $F_L^{c,b}$
is viable, and the NNLO predictions are stable, exhibit reduced scale
dependence, and will minimize the ``tuning'' of both the $m_c$ and the
$\xi$-scaling parameter.
S-ACOT-$\chi$ is the default heavy-quark scheme of CTEQ PDF analyses
and it will play a significant role in global fits at NNLO.


S. Alekhin (in collaboration with S. Moch) presented the a new
calculation which implemented the running mass definition for the DIS
semi-inclusive structure function.\cite{alekhin}  
The goal of this approach is to
obtain improved perturbative stability with reduced scale variation,
and a better determination of the heavy quark PDFs.  Comparisons with
the HERA data for $F_2^{c\bar{c}}$ are presented and yield good
agreement.  Additionally, the extracted value of the charm mass
compares favorably with the PDG value.


R. Placakyte (on behalf of the H1 and ZEUS collaborations) presented a
study of the combined H1 and ZEUS $F_2^{c\bar{c}}$ data investigating
the effects of the renormalization scheme and the charm mass.\cite{placakyte}  
By
treating the charm mass as a free parameter and allowing the fit to
select an optimal value, they discovered that different schemes yield
very different values for the charm mass.  They then studied the
impact of these differences for W production at the LHC.  Remarkably,
they find that if the schemes and charm mass values are implemented
consistently, the net effect on the predicted W boson production cross
section is largely compensated so that the total uncertainty is
roughly 1\%; in contrast, the uncertainty associated with a
straightforward variation of the charm mass can be as much as 5\%.
Thus, the use of the optimal charm mass serves to reduce the
uncertainty of the cross section predictions.


J. Rojo (on behalf of the NNPDF collaboration) investigated the impact
of the heavy quark mass on the PDFs.\cite{rojo}  
Specifically, he presented a new
fit (NNPDF2.1) which was performed in a General-Mass
Variable-Flavor-Number (GM-VFN) scheme, and compared this to the
Zero-Mass result (NNPDF2.0).  The two sets of PDFs were consistent at
the 1$\sigma$-level, and most of the differences arose at medium to
small $x$ values.  To see the impact on physical observables, they
studied the sensitivity of the W and Z cross sections; for charm mass
variations in the range [1.4, 1.7]~GeV, they find the variation of the
cross section is generally at the 1$\sigma$-level or less.  The
variation of the b-quark mass could be important for b-quark initiated
processes such as $b\bar{b}\to H$ or t-channel single-top
production. Figure~\ref{fig:rojoTZ}-a) displays the relative
uncertainty of the Higgs production cross section combining the PDF
uncertainty with the corresponding uncertainty on the b-quark mass;
this can grow as large as 10\% for large Higgs mass values.


K. Kovarik (in collaboration with T. Stavreva, I. Schienbein,
F. Arleo, F. Olness, J. Owens, J.Y. Yu) presented new results on
direct photon production in association with heavy quarks.\cite{Kovarik:2011br} 
This
process can be used to constrain both the heavy quark and the gluon
PDF.  The $\gamma c$ and $\gamma b$ final states have been studied at
the Tevatron; preliminary measurements indicate the bottom channel
appears to agree well with predictions, while there is an apparent
discrepancy in the charm process at high $p_T$.  Curiously, this is
what one might expect if there were an intrinsic charm component in
the PDFs. Figure~\ref{fig:rojoTZ}-b) displays the predictions 
for $p\bar{p}\to
\gamma c$ without any intrinsic charm, and with two intrinsic charm
(IC) models (BHPS IC and Sea-like IC); the intrinsic charm PDF
component appears to yield better agreement between data and the
predictions.  A more complete analysis of the Tevatron data is in
progress, and studies involving LHC data will soon be available.
Additionally, this process has the potential to be one of few processes that can
help constrain the largely unconstrained nuclear gluon PDF in pA
collisions at the LHC.


A. W. Jung (on behalf of the H1 collaboration) presented new results
on $D^*$ production in DIS at low $Q^2$.  This analysis increased the
phase space in $\eta$ and $p_T$ of the measurements, and compared
these results to different theoretical predictions.\cite{Jung:2011iu}
Specifically, the
Zero Mass (ZM) calculation failed to adequately describe the data
while the massive HVQDIS program (based upon the $g\to Q \bar{Q}$
process) generated a reasonable description. With this expanded $\eta$
and $p_T$ there were some regions at the limit of the kinematics where
the HVQDIS model differed from the data. It will be interesting to see
if an improved theoretical analysis can yield improved comparisons in
this region.
This differential measurement is also used for determining the charm
contribution to the proton structure. As this measurement yields
largest phase space coverage, and thus smallest errors at low $Q^2$
for this kind of measurement, it has a strong impact on the HERA
combined $F_2^{c\bar{c}}$ extractions.



\section{Experimental Issues:}  

\subsection{Open Charm and Beauty Production, Quarkonia
and spectroscopy }
At this conference we have seen a wealth of new results
on charm and beauty quark production.
For the first time results in hadroproduction are available from LHC
but also new DIS and photoproduction results were released from HERA.
In general the description of all these measurements by the 
massive scheme NLO predictions is reasonable, with some
exceptions in certain phase spaces corners.
New TEVATRON results were presented on quarkonia,
there are still both experimental and theoretical
puzzles, in particular in the area of polarisation.
BELLE and BABAR showed new interesting spectroscopy results
for quarkonia and on excited charm mesons.
The results mentioned in the following  subsection are only (hopefully)
representative samples from selected talks.
Many more interesting results can be found in these 
and the other working group session talks
to which no explicit reference is given here.
\subsection{Open Charm and Beauty Production}
\begin{figure}
  \includegraphics[width=.49\textwidth]{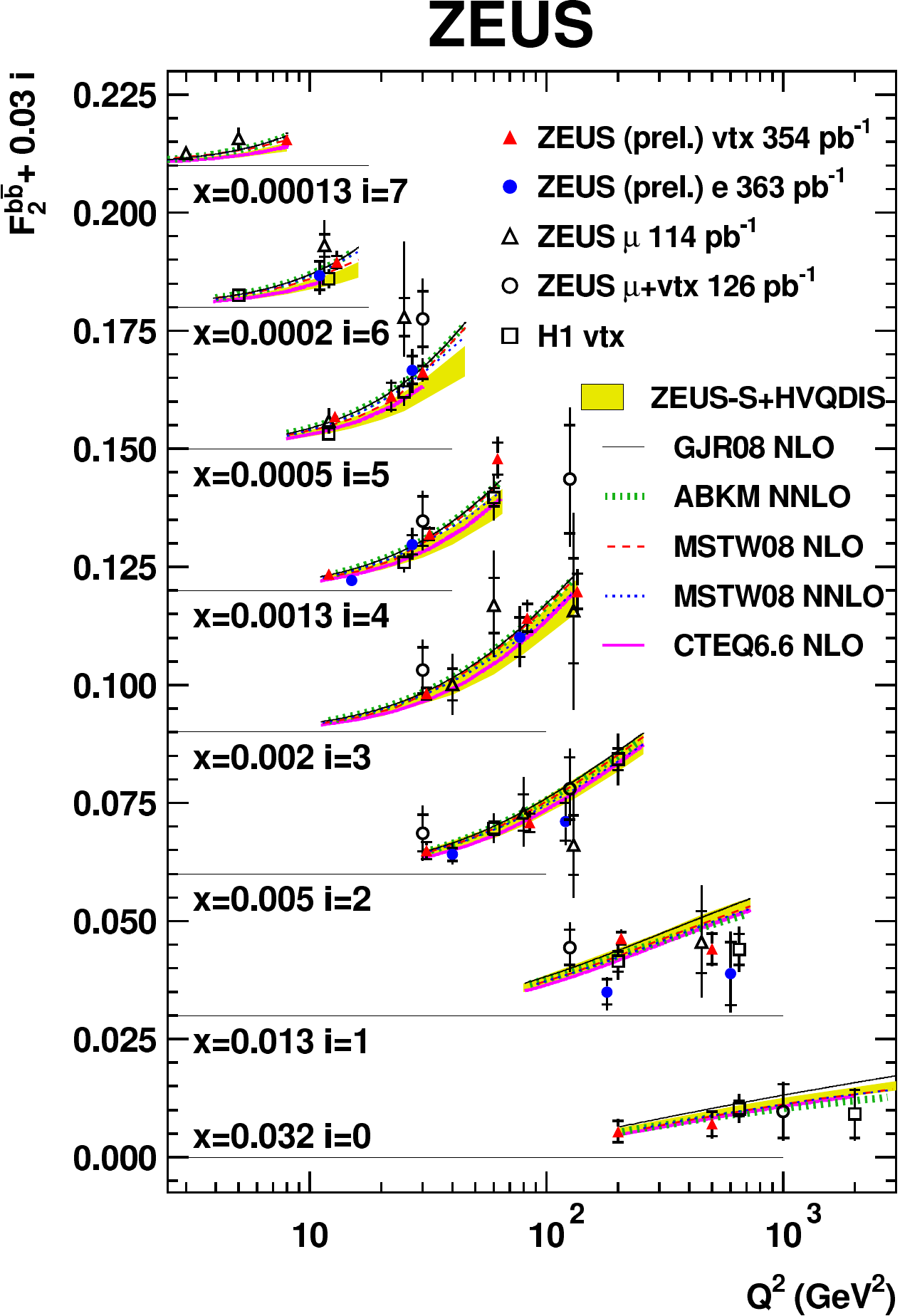}
  \includegraphics[width=.49\textwidth]{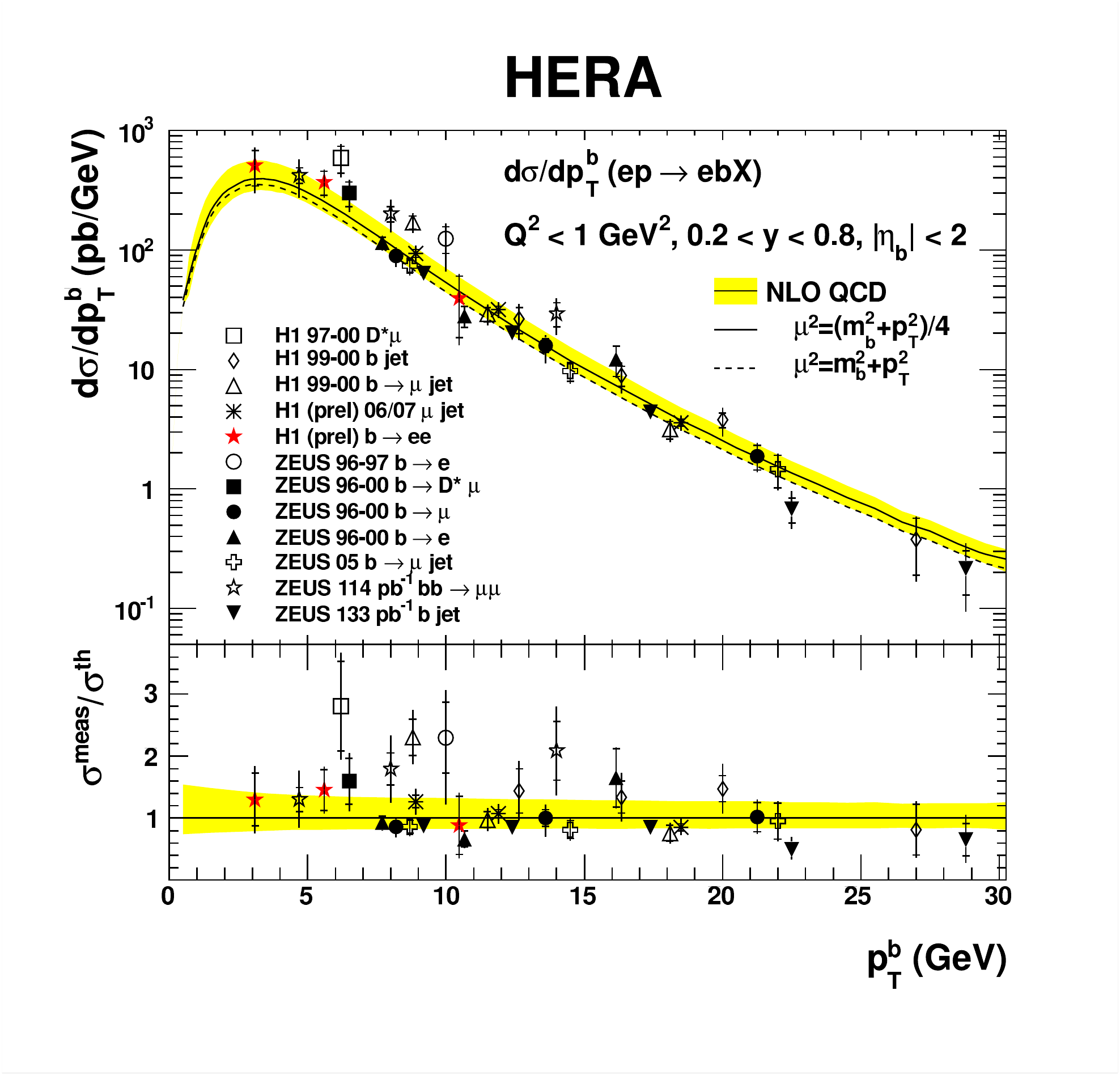}
  \caption{Left: Compilation of $F_2^{b\bar{b}}$ results from H1 and ZEUS compared
           to various predictions. Right: Comparison of H1 and ZEUS beauty
measurements in photoproduction to a NLO prediction.}
\label{fig:f2bb}
\end{figure}

\begin{figure}
  \includegraphics[angle=-90,width=.7\textwidth]{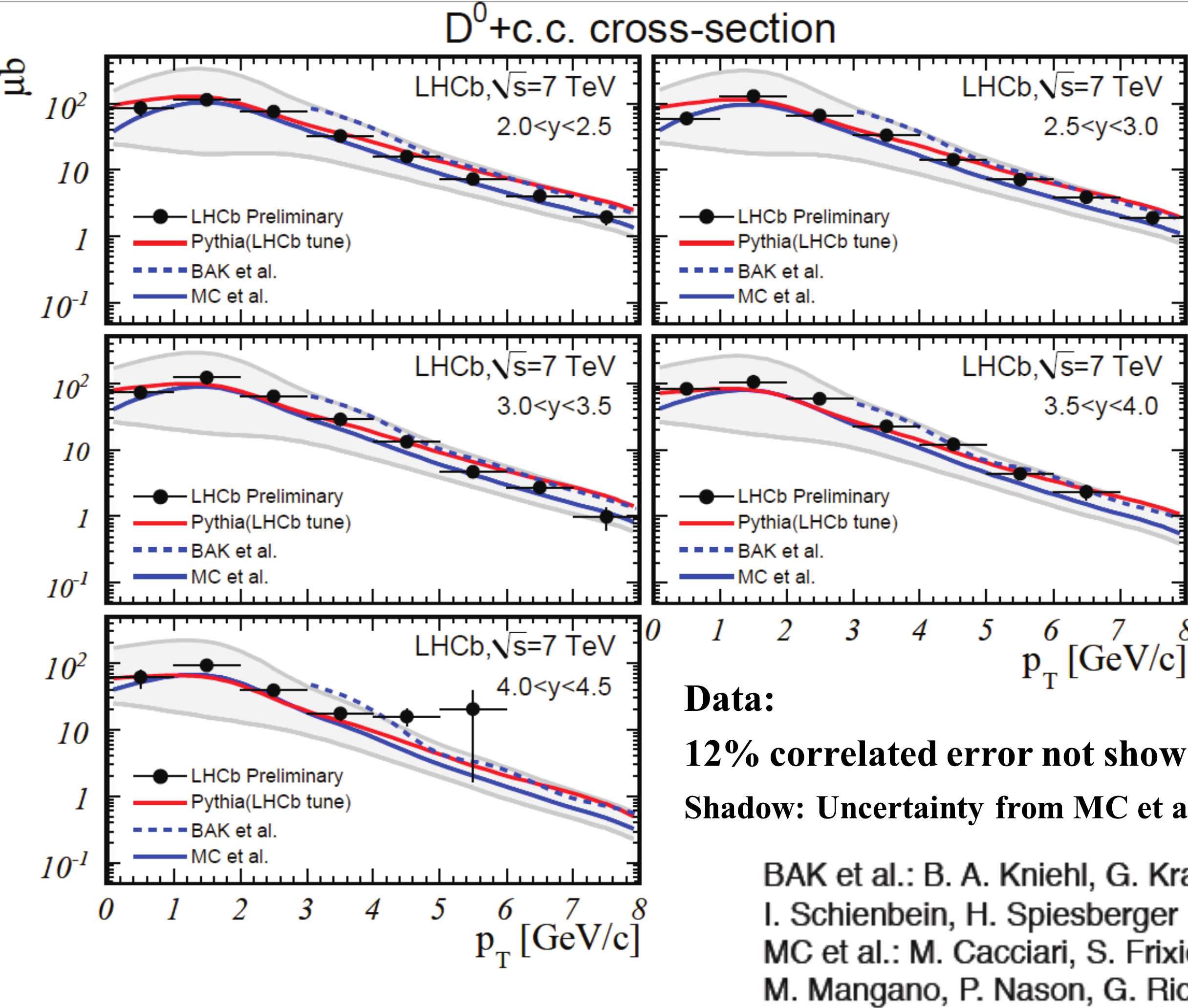}
  \caption{$D^0$ production results from LHCb.}
  \label{lhcbd0}
\end{figure}

\begin{figure}
\unitlength1cm
\begin{picture}(5,7)
  \put(-5.,7.){\includegraphics[angle=-90,width=.5\textwidth]{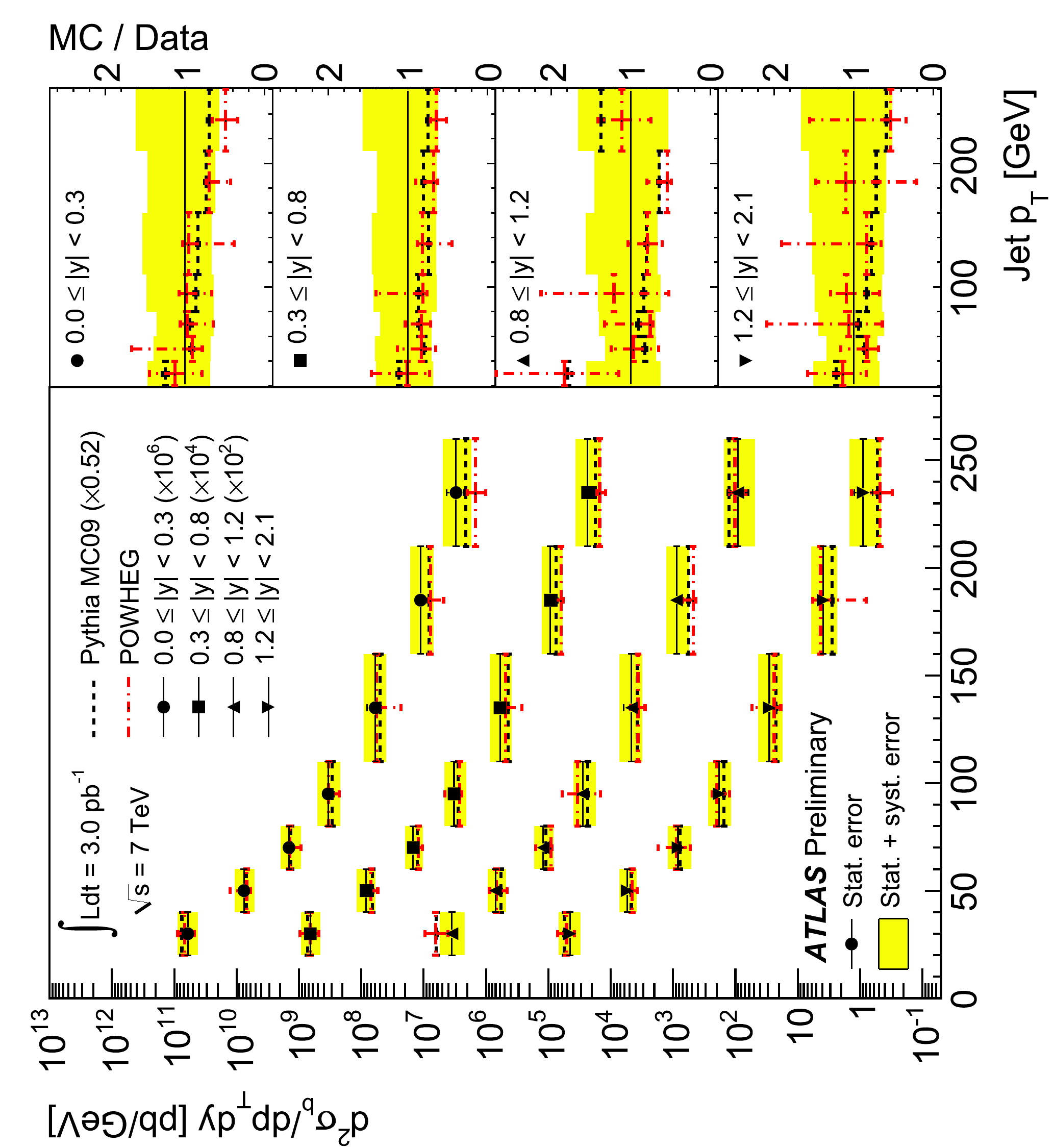}}
  \put(2.6,0.){\includegraphics[width=.47\textwidth]{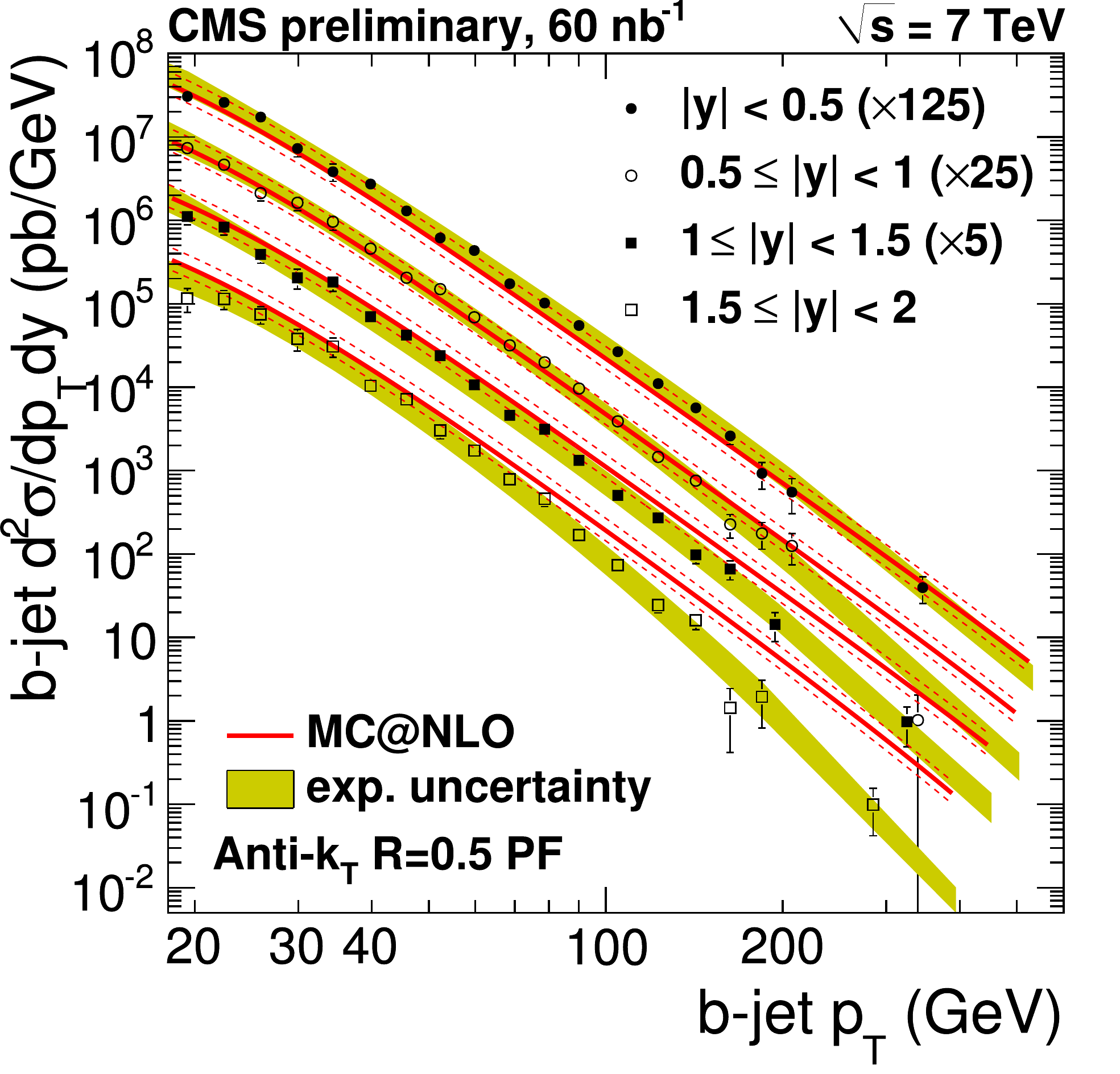}}
\end{picture}
  \caption{Beauty jet production measurements by ATLAS and CMS.}
  \label{fig:blhc}
\end{figure}

\paragraph{Production in DIS at HERA:}
New measurements on the structure function $F_2^{b\bar{b}}$ were presented
by ZEUS \cite{Shehzadi:2011vy}, \cite{slava}, using electrons in the final state
or based on a secondary vertex tag. 
The compilation of all available HERA results 
is shown in Fig. \ref{fig:f2bb} (left).
The description by the various theory predictions is reasonable.
\paragraph{Production in Photoproduction at HERA:}
Figure. \ref{fig:f2bb} (right)  
shows an updated summary plot 
for the beauty production measurements
in photoproduction at HERA as function of the 
beauty quark transverse momentum.
Two new measurements were included:
The first one from H1\cite{sauter} extends the range to
lower transverse momenta (down to about 2 GeV) 
than ever covered before at HERA.
The second result from ZEUS\cite{brock} adds precise points at
large momenta (up to 30 GeV).
The summary plot demonstrates that the massive
scheme NLO calculation can describe the HERA data
over the whole kinematic phase space reasonably well.
\paragraph{Charm and beauty hadroproduction at LHC:}
Despite the low luminosities
recorded and analysed at the time of this conference,
an impressive amount of charm and beauty 
hadroproduction results was already presented
by the LHC experiments.
ATLAS \cite{Chiodini} showed first charm meson production results
which were compared with
different NLO calculations which are able to 
describe the data within the theory uncertainties.
So far the data are restricted to transverse momenta
up to 40 GeV, but with the much more data already
recorded one can expect in the future very interesting extensions
to much higher momenta.
LHCb \cite{zhang} showed (Fig. \ref{lhcbd0}) differential cross sections for
$D^0$ production versus transverse momentum 
in several bins of rapidity, in the typical
forward rapidity range (2-4.5) covered by the apparatus.
Both ATLAS \cite{ruiz} and CMS \cite{schmidt1,schmidt2} have presented measurements
of beauty jet production 
as function of the jet transverse momentum
in bins of rapidity (see Fig. \ref{fig:blhc}).
Again the description by NLO calculations is reasonable,
at high $p_T$ the MC@NLO predictions overshoot the CMS data
in the not so central rapidity ranges.
CMS \cite{rizzi} has also studied beauty-beauty angular correlations.
The gluon splitting $g\rightarrow b\bar{b}$ is expected
to give the dominant contribution.
The description by different models is not perfect.
ALICE \cite{bianchin} has presented measurements of charm plus beauty 
cross sections tagged with muons measured in the
ALICE forward muon spectrometer.
Also here the description by the prediction (FONLL) 
is fairly good.
%
%
%
The future of all the LHC open charm and beauty measurements
is bright: more statistics and in particular kinematic phase-space extensions
to much higher scales will be possible.
%
%

\subsection{Quarkonia production}
In this subsection we will review talks from
LHC, TEVATRON and HERA on inelastic $J/\Psi$ production.
Quarkonia production at hadron colliders is theoretically
not settled.
Most models fail to simultaneously describe 
measurements of both cross section 
and polarisation.
Prompt $J/\psi$ cross section measurements have
been presented by CMS \cite{ronchese}. 
They exceed expectations in the forward region
at small $p_T$.
Measurements of Upsilon(1s) production from
LHCb and CMS were presented in the talk \cite{anderson}.
LHCb provides a nice extension to the forward regions.
A Good agreement at high $p_T$ between the CMS data and
NRQCD (non relativistic QCD) prediction was reported.
ZEUS \cite{alessandro} presented $J/\psi$ photoproduction cross sections
at HERA, double differential as function of $p_T$
and the energy fraction $z$ which the $J/\psi$ takes
from the photon in the proton rest frame.
These data will provide important input for global fits 
with the NRQCD approach.
Upsilon(1S) polarisation measurements from CDF and
D0 are largely inconsistent, as presented in the talk \cite{gerberich}.
%
%
However, one difference is that they measure in different rapidity regions.
%
The CDF $J/\psi$ polarisation measurements are not well described
by neither NRQCD nor kT-factorisation calculations.
In summary, there are still both experimental and theoretical
puzzles in the area of quarkonia polarisation.
\subsection{Charm and Beauty Spectroscopy}
Charmonium spectroscopy is an exciting topic
as pointed out in the talk \cite{rakitin}.
There are several states, recently discovered,
that are  non-compliant with the $q\bar{q}$ hypothesis.
An Example for this is the $X(3872)$ state.
Various ideas exist for exotic charmonium-like states, e.g.
tetraquark models, hybrids and others.
LHCb measurements of
the masses of $X(3872)$ and for beauty ground states
mesons/baryons were presented in the talk \cite{cian}, 
the latter one being already the world best.
The $h_b$ members of the bottomonium 
family were searched for by Belle as reported
in the talk \cite{piilonen}.
Prominent peaks are observed and 
mass values were determined for the 1P and 2P states.
BABAR results on $h_b$ were reported by \cite{rakitin}
and the mass value of the 1p state was measured.
As summarised by Leo Piilonen: B factories have made surprising
studies of new states in charmonia/bottomonia, but more
studies and more data are needed to reveal their true nature.
New results have been also presented on the 
excited charm meson spectroscopy, by
BABAR \cite{cibinetto} and ZEUS \cite{andrii}.
In these studies BABAR has discovered new states
which are attributed to be radial excitations.
%
%



\subsection{Heavy Flavor in Heavy Ion Collisions}

At this conference, the two RHIC collaborations, STAR and PHENIX, and the ALICE
collaboration at the LHC, presented measurements of multi-particle
correlations involving a heavy flavor parent as well as measurements pertaining
to the interaction of $J/\psi$ and $\Upsilon$ mesons in the medium produced in
heavy ion collisions.

\begin{figure}

\includegraphics[width=0.50\textwidth]{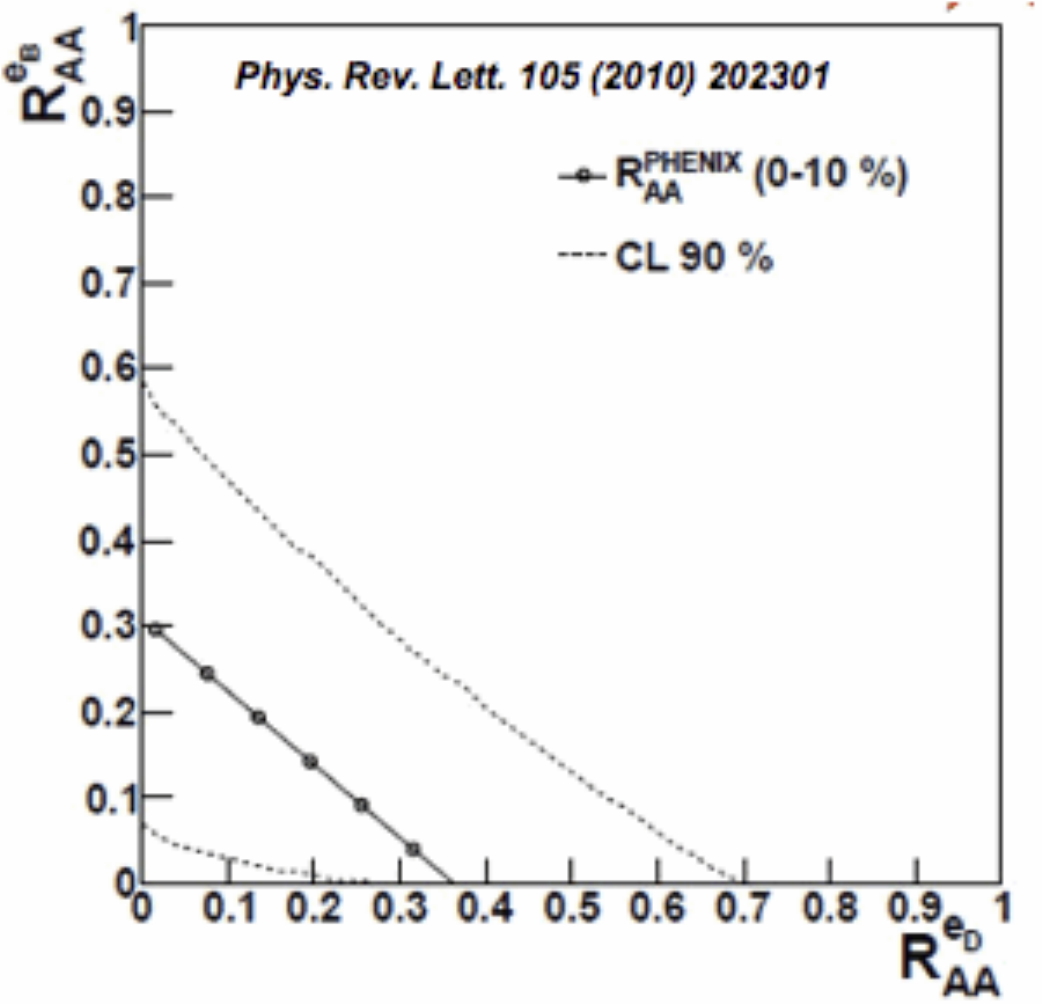}
\caption{90\% confidence level contours for the nuclear suppression factor of
open B and D mesons, using non-photonic electron $R_{AA}$ measurements from
PHENIX and electron-hadron correlations form STAR.}
\label{fig:STAR_RAA}

\end{figure}

PHENIX \cite{Engelmore} showed two-particle correlations
of electron-muon pairs in d+Au and $p+p$ collisions coming predominantly from
decays of open charm mesons.  Back-to-back $e-\mu$ pairs are suppressed in d+Au
collisions relative to $p+p$ collsions.  Along these same lines, the STAR
experiment \cite{X_Li} presented a measurement of correlations between an
electrons from charm and bottom decays and a hadron in d+Au and $p+p$
collsions.  The back-to-back $e-\textrm{h}$ correlation as a function of the
azimuthal angle between the pairs is seen to be broadened with respect to the
distribution in $p+p$, similar to that seen in di-hadron correlations.  Using
$e-\textrm{h}$ correlations as well as data from the PHENIX experiment, STAR was
also able to put an upper limit of about 0.6 with 90\% confidence level on the
nuclear suppression factor $R_{AA}$ of open bottom mesons in Au+Au collisions at
$\sqrt{s_{NN}}$ = 200 GeV.  Figure \ref{fig:STAR_RAA} shows the 90\% confidence
level for the D and B meson suppression.

The PHENIX collaboration \cite{K_Lee} presented measurements of the suppression
of $\Upsilon$ mesons at low $x$ in the di-muon channel, as well as high
precision measurements of $R_{AA}$ for $J/\psi$ mesons at forward at
mid-rapidity.  STAR \cite{Cervantes} showed a proof-of-principle result on a
correlation between $\Upsilon$ mesons and unidentified hadrons, which could be
used to measure the radiation
emitted off of a colored heavy quark pair during
production.  The ALICE experiment \cite{Gagliardi:2011hf} presented early results on
$J/\psi$ production through the di-electron and di-muon channels.

\subsection{Top Quark Production}

CDF, D0, ATLAS, and CMS presented improved measurements of the top quark mass
and production cross sections.  Results on top quark charge and asymmetry
between $t$ and $\bar{t}$ masses were also presented.

The CDF collaboration \cite{Moon:2011vv} excluded the top quark charge of -4/3 with a
95\% confidence level.  This presentation also showed an initial look at W
boson polarization, measured to be consistent with standard model predictions.
D0 \cite{Gerber} presented the most precise result in the dilepton
channel of the top quark cross section $\sigma_{t \bar{t}}$, which was measured
to be $7.4^{+0.9}_{-0.8}$ pb.  D0 also measured the t-channel cross section to
be consistent with the standard model.  The same collaboration also presented
their most precise top quark mass measurement to data \cite{Jung}, as well as a
measurement, consistent  with zero, of the mass difference between the top and
anti-top quarks.  

\begin{figure}

\includegraphics[width=.55\textwidth]{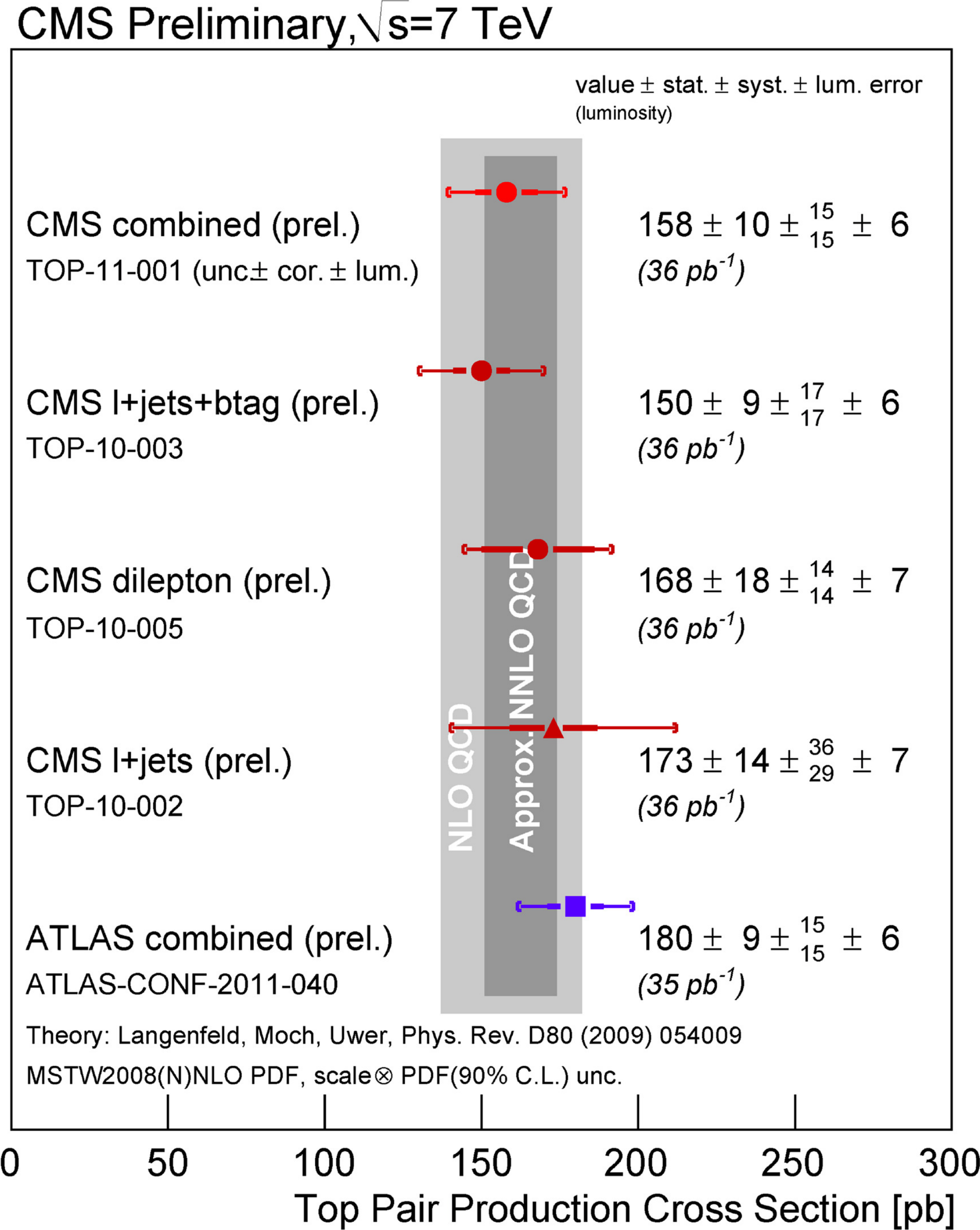}
\caption{ATLAS and CMS measurements of the top quark cross section.}
\label{fig:LHC_top}

\end{figure}

The ATLAS collaboration presented a measurement of the top quark cross section
at 7 TeV by tagging a $b$-jet and measuring 1-lepton and 2-lepton rates
\cite{Okumura:2011ex}, and also showed a measurement of the top quark mass using 3
complementary techniques \cite{Fiolhais}.  CMS \cite{Caudron} showed
measurements of the top quark cross section from 3 methods, as well as a
measurement of the top quark mass in the dilepton channel \cite{Silva}. Figure
\ref{fig:LHC_top} shows the cross section measurements from LHC and ATLAS. CMS
\cite{Klingebiel} also showed t-channel cross section results with 36\%
precision, consistent with the standard model.

\subsection{Rare Searches}

D0 \cite{Badaud}, CDF \cite{Chen}, ATLAS \cite{Lenz}, and CMS \cite{Pagano,
Anghel} showed many results of new physics searches involving top quarks.  The
closest any of the results came to showing an inconsistency with the standard
model was the measurement by CDF of the $t\bar{t}$ rest frame asymmetry in 2
$\sigma$ excess over the standard model prediction.  Other results set limits
on masses of new particles, such as the limit set by D0 on the mass of a fourth
generation quark to be below 285 GeV/$c^2$ with a 95\% confidence level.


\begin{theacknowledgments}

We thank the presenters of the Heavy Flavours Session 
for their contributions. 
We also thank the convenors of 
{\it WG1: Structure Functions and Parton Densities}
and 
{\it WG3: Electroweak Physics and Beyond the Standard Model}
for coordinating joint sessions with our session. 
%
%
This work was partially supported by the U.S.\ Department of Energy
under grant DE-FG02-04ER41299,
and the Lightner-Sams Foundation.

\end{theacknowledgments}
\bibliographystyle{utphys}
\bibliography{hqSummary}

\end{document}